\documentclass{phb-proc4-auth}

\usepackage{graphicx}
\usepackage{amssymb}

\newcommand{\gsim}{\raisebox{-3pt}{$\stackrel{>}{\sim}$}}
\newcommand{\n}{\langle n \rangle}

\begin{document}
\begin{frontmatter}

\journal{SCES '04}

\title{Finite-temperature investigation of quarter filled ladder systems}

\author[TU]{C.~Gabriel},
\author[TU]{E.~Sherman},
\author[TU]{T.~C.~Lang},
\author[TU]{M.~Aichhorn},
\author[TU]{H.~G.~Evertz\corauthref{1}}

\address[TU]{Institute for Theoretical and Computational Physics, Graz
  University of Technology, Austria}

\corauth[1]{Corresponding author: e-mail: evertz@tugraz.at}

\begin{abstract}
We investigate charge ordering in a quarter-filled ladder at finite temperature
by determinantal Quantum Monte Carlo.
The sign problem is moderate in a wide range of model parameters 
relevant for NaV$_2$O$_5$.
The charge order parameter exhibits a crossover as a function of 
inverse temperature $\beta$ on finite systems.
Above a critical nearest neighbor Coulomb repulsion $V_{\rm c}$,
the correlation length grows exponentially with $\beta$,
indicative of the ordered phase at $\beta=\infty$. 
We find a clear single-particle gap manifesting itself in a flat
$n(\mu)$ dependence at large nearest neighbor Coulomb repulsion $V$.
\end{abstract}


\end{frontmatter}

The inorganic ladder compound NaV$_2$O$_5$ has attracted great attention in
recent years. This interest was triggered by magnetic susceptibility
measurements \cite{Isobe96}, which show a phase transition at $T=34$\,K into a
low-temperature spin-gapped phase. This transition is accompanied by charge
ordering, as observed in NMR measurements \cite{Ohama99}, where the valence of
the vanadium sites changes from V$^{4.5}$ to V$^{4.5\pm\delta}$, with $\delta$
the amount of charge disproportion. This transition has been studied theoretically by
several techniques at $T=0$ \cite{t0}.

On a microscopic level the system can be described by an extended Hubbard model 
\begin{eqnarray}
  H=&-&\sum_{\langle ij\rangle,\sigma}t_{ij}(c_{i\sigma}^\dagger
  c_{j\sigma}^{\phantom{\dagger}}+\mbox{H.c.})+U\sum_in_{i\uparrow}n_{i\downarrow}\nonumber\\
  &+&V\sum_{\langle ij\rangle}n_in_j -\mu\sum_in_i,\label{eq:ham}
\end{eqnarray}
at quarter filling $\n=0.5$,
with hopping matrix elements $t_{ij}=t_x $ along the ladder 
and $t_{ij}=t_y $ within a rung,
and chemical potential $\mu$.
We state all energies in units of $t_y$. 
These hopping parameters as well as the onsite
Coulomb interaction can be extracted from first-principle calculations \cite{LDA}.
The hopping along chains  $t_x\simeq 0.5 t_y$ is weaker than along rungs.
This strongly influences the physics of the ladder, 
for which a spin-gap seems to appear at $t_x \gsim t_y$ \cite{t0}.

We used $t_x=0.5$ and $U=8$. Since the non-local Coulomb
interaction $V$ cannot be determined properly by band-structure calculations,
we used $V$ as a free parameter of the Hamiltonian. 
The charge order parameter is 
 $\Delta_{\rm  co}^2 = \frac{1}{2L\n}
                       \sum_{ij}e^{{\rm i}{\mathbf Q}({\mathbf r}_i-{\mathbf r}_j)}
                       (n_i-\n)(n_j-\n)$ 
with ${\mathbf Q}=(\pi,\pi)$,
which is unity for complete ordering.

\begin{figure}[t]
  \centering
  \includegraphics[width=0.7\columnwidth]{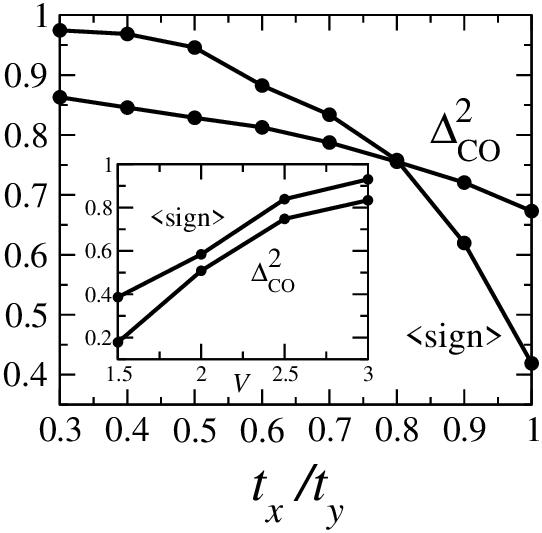}
  \caption{\label{sign_and_CO}%
    Mean value of sign and order parameter $\Delta_{\rm co}^2$ as functions of
    $t_x/t_y$ at $V=3$, $\beta=6$, $L=16$, $\n =0.5$,
    and as functions of $V$ at $\beta=6$ (inset).} 
\end{figure}

We performed grand canonical calculations by  determinantal quantum Monte Carlo.
These are often very difficult for doped systems because of a sign problem.
Fortunately, the average sign is favorably large in the relevant parameter range
of $t_x/t_y=0.5$ and large $V$ (Fig.~\ref{sign_and_CO}).
In the opposite case of isotropic $t_x=t_y$ at small $V$, $\langle {\rm sign}\rangle$
becomes very small.
The charge order parameter exhibits similar behavior,
but it is less strongly dependent on $t_x/t_y$.
Charge order grows with increasing $V$.

\begin{figure}[t]
  \centering
  \includegraphics[width=0.72\columnwidth]{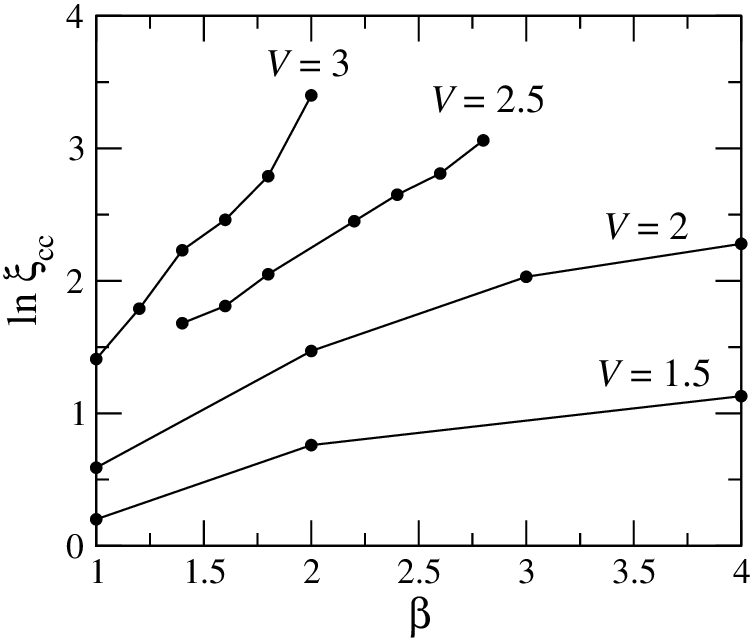}
  \caption{\label{xi}%
    Logarithm of the charge correlation length $\xi_{\rm cc}$ 
    as a function of $\beta$ for different interactions $V$
    ($L=32$ for $V=1.5$ and $2$; $L=44$ for $V=2.5$ and $3$)}
\end{figure}

Fig.~\ref{xi} shows the charge correlation length $\xi_{\rm cc}$.
At small interactions, $V=1.5$ and $2.0$, the correlation length  seems to saturate,
but for $V=2.5$ and $3.0$ it increases exponentially with $\beta$,
with a $V$-dependent slope.
This behavior is consistent with the  
1D Ising model in a transverse field (IMTF) \cite{imtf}, which
is equivalent to 
Eq.~\ref{eq:ham} in the limit of one spinless electron per rung. For large
$V$, the transverse field goes to zero, and 
$\xi_{\rm IMTF}=|\ln\tanh(\beta)|^{-1}$.
This is exponential behavior with slope 2 at large $\beta$, 
which the data in Fig.~\ref{xi} appear to approach.
There is long range order in the 
thermodynamic limit only at $\beta = \infty$. 
For weaker interactions, $V<2t_y$, 
the correlation length $\xi_{\rm IMTF}$ remains finite even in the
limit $\beta\to\infty$, showing a disordered phase at all temperatures.
The results in Fig.~\ref{xi} are nicely consistent with
recent DMRG calculations \cite{DMRG} which show that at $T=0$ the system has a
quantum phase transition to an ordered phase at $V_c=2.1(1)$.

\begin{figure}[t]
  \centering
  \includegraphics[width=0.72\columnwidth]{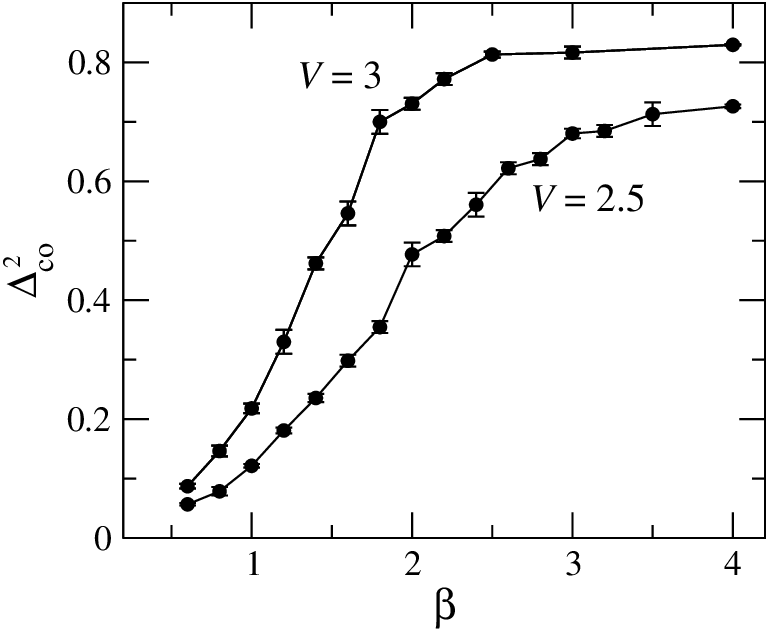}
  \caption{\label{CO_of_beta}%
    Charge order parameter as a function of $\beta$ for $V=3.0$ and $V=2.5$
    with $L=32$ rungs.} 
\end{figure}

However, the behavior of finite size systems is different from the IMTF in
the thermodynamic limit.
As a function of inverse temperature, the charge order parameter exhibits a
crossover at large $V$ (Fig.~\ref{CO_of_beta}).
For the single ladder, this crossover is a finite size effect. 
It appears since at some $\beta$, the charge order correlation length will exceed the system size, 
resulting in apparent long range order. 
For a three-dimensional system of coupled ladders, this crossover can become a phase transition.
The order parameter at $V=3$ and  different $\beta$ scales well as a function of $\xi(L)/L$.

\begin{figure}[t]
  \centering
  \includegraphics[width=0.7\columnwidth]{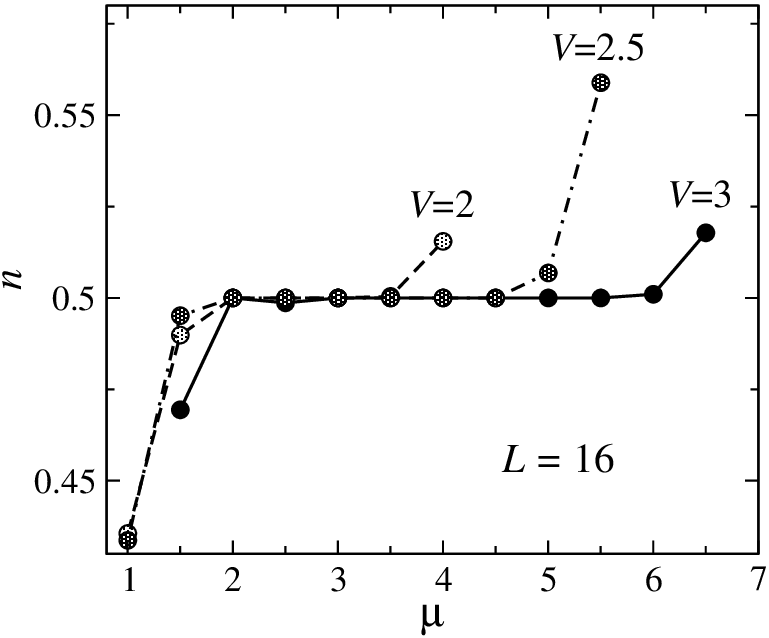}
  \caption{\label{fig_single_particle_gap}
    Mean electron density  as a function of the chemical potential
    $\mu$ at $\beta=6$, L=16.}
\end{figure}

The onset of charge order at large $V$ is most clearly visible in the single particle gap
shown in Fig.~\ref{fig_single_particle_gap}.
It manifests itself
as a plateau in the $n(\mu)$ dependence where $\n=0.5$ remains constant in a
region $\mu_{\rm min}<\mu <\mu_{\rm max}$. At large $V$, the upper boundary
$\mu_{\rm max}$ shifts with $V$ as approximately $3V$, which is the same value
as in the atomic limit at full ordering.  

We greatfully acknowledge support by the Austrian Science Fund (FWF), project
P15520. M.A. is supported by DOC (Austrian Academy of Sciences).

\end{document}